\documentclass[aps,prd,superscriptaddress,showkeys,nofootinbib]{revtex4-2}
\usepackage[english]{babel}
\usepackage{amsmath,bm}
\usepackage{amssymb}
\usepackage{enumitem}
\usepackage{textcomp}
\usepackage{graphicx}
\usepackage{dsfont}
\usepackage{color}

\definecolor{GREEN}{rgb}{0.,0.5,0}
\definecolor{BLUE}{rgb}{0.,0.,0.75}

\usepackage[breaklinks=true]{hyperref}
\hypersetup{
	colorlinks=true,
	linkcolor=BLUE,
	filecolor=magenta,
	urlcolor=BLUE,
	citecolor=BLUE,
}

\usepackage{mathrsfs}
\usepackage[normalem]{ulem}

\usepackage{placeins}

\begin{document}

	
	\title{Chirality production during axion inflation}
	
	\author{E.V.~Gorbar}
	\affiliation{Physics Faculty, Taras Shevchenko National University of Kyiv, 64/13, Volodymyrska Str., 01601 Kyiv, Ukraine}
	\affiliation{Bogolyubov Institute for Theoretical Physics, 14-b, Metrologichna Str., 03143 Kyiv, Ukraine}
	
	\author{A.I.~Momot}
	\affiliation{Physics Faculty, Taras Shevchenko National University of Kyiv, 64/13, Volodymyrska Str., 01601 Kyiv, Ukraine}
	
	\author{I.V.~Rudenok}
	\affiliation{Physics Faculty, Taras Shevchenko National University of Kyiv, 64/13, Volodymyrska Str., 01601 Kyiv, Ukraine}
	
	\author{O.O.~Sobol}
	\email{oleksandr.sobol@epfl.ch}
	\affiliation{Institute of Physics, Laboratory of Particle Physics and Cosmology, \'{E}cole Polytechnique F\'{e}d\'{e}rale de Lausanne, CH-1015 Lausanne, Switzerland}
	\affiliation{Physics Faculty, Taras Shevchenko National University of Kyiv, 64/13, Volodymyrska Str., 01601 Kyiv, Ukraine}
	
	\author{S.I.~Vilchinskii}
	\affiliation{Physics Faculty, Taras Shevchenko National University of Kyiv, 64/13, Volodymyrska Str., 01601 Kyiv, Ukraine}
	\affiliation{D\'{e}partement de Physique Th\'{e}orique, Center for Astroparticle Physics, Universit\'{e} de Gen\`{e}ve, 1211 Gen\`{e}ve 4,  Switzerland}

	\date{\today}
	
\begin{abstract}
    We study the generation of the chiral charge during axion inflation where the pseudoscalar inflaton field $\phi$ couples axially to the electromagnetic field through the term $(\beta/M_p)\phi\,\boldsymbol{E}\cdot\boldsymbol{B}$ with dimensionless coupling constant $\beta$. To describe the evolution of electromagnetic field and determine $\langle\boldsymbol{E}\cdot\boldsymbol{B}\rangle$ sourcing the chiral asymmetry during inflation due to the chiral anomaly, we employ the gradient expansion formalism. It operates with a set of vacuum expectation values of bilinear electromagnetic functions and allows us to take into account the backreaction of generated fields on the inflaton evolution as well as the Schwinger production of charged fermions. In addition, we include the chiral magnetic effect contribution to the electric current $\boldsymbol{j}_{\rm CME}=e^{2}/(2\pi^2)\mu_{5}\boldsymbol{B}$, where $\mu_5$ is the chiral chemical potential which quantifies the chiral charge production. Solving a set of equations for the inflaton field, scale factor, quadratic functions of the electromagnetic field, and the chiral charge density (chiral chemical potential), we find that the chirality production is quite efficient leading to the generation of a large chemical potential at the end of axion inflation.
\end{abstract}

	\maketitle

	\section{Introduction}
	\label{sec-intro}

The inflation paradigm is a very successful idea which solves a large number of cosmological problems. In particular, it provides the mechanism for the generation of primordial scalar and tensor perturbations which then imprint in the cosmic microwave background spectrum and lead to the large-scale structure formation \cite{Harrison:1970,Zeldovich:1972,Chibisov:1982} (see Refs.~\cite{Mukhanov:1992,Durrer:book} for a review). Moreover, it is often assumed that the primordial magnetic field originates from inflation as well (see the seminal papers in Refs.~\cite{Turner:1988,Ratra:1992,Garretson:1992,Dolgov:1993}). This field then may serve as a ``seed'' for magnetic fields observed in astrophysical objects such as galaxies and galaxy clusters \cite{Grasso:2001,Kronberg:1994,Widrow:2002,Giovannini:2004,Kandus:2011,Vallee:2011,Ryu:2012,Durrer:2013,Subramanian:2016}. On the other hand, in cosmic voids containing vanishingly small amounts of matter, it may still remain in the form unchanged by astrophysical processes and, thus, contain very important information about earlier stages of the Universe's history. Indeed, there are evidences for the presence of magnetic fields in voids following from the gamma-ray observations of distant blazars \cite{Tavecchio:2010,Ando:2010,Neronov:2010,Tavecchio:2011,Dolag:2010,Dermer:2011,Taylor:2011,Huan:2011,Vovk:2012,Caprini:2015} (for a recent review, see Ref.~\cite{Batista:2021}).

One of the simplest and widely accepted inflation models involves a real scalar field $\phi$, dubbed inflaton, which slowly rolls along the slope of its effective potential and exhibits the vacuumlike equation of state $p=-\rho$. However, for this regime to occur one needs a sufficiently flat inflaton potential. On the other hand, coupling of the inflaton to matter fields (which is needed for successful reheating of the Universe after inflation) may lead to quantum corrections to the inflaton potential which can spoil its flatness and disable inflation. Therefore, one needs a mechanism to preserve the form of the inflaton potential at least in the range of the inflaton values relevant for inflation. One example of such a mechanism is realized in axion inflation where the role of the inflaton is played by the pseudo-Nambu-Goldstone boson of the shift symmetry -- the axion field. In application to magnetogenesis, the axion inflation is also very advantageous because it produces helical magnetic fields in the early Universe \cite{Anber:2006,Anber:2010,Durrer:2011,Barnaby:2012,Caprini:2014,Anber:2015,Ng:2015,Fujita:2015,Adshead:2015,Adshead:2016,Notari:2016,Jimenez:2017cdr,Domcke:2018,Cuissa:2018,Shtanov:2019,Shtanov:2019b,Sobol:2019,Domcke:2019bar,Domcke:2019,Domcke:2020,Gorbar:2021}. This property greatly increases the chances for survival for these magnetic fields compared to the case of nonhelical magnetic fields due to the inverse cascade of magnetic helicity \cite{Joyce:1997uy,Boyarsky:2011uy,Banerjee:2004,Tashiro:2012mf,Hirono:2015rla,Dvornikov:2016jth,Gorbar:2016klv,Brandenburg:2017rcb,Schober:2018wlo}.

In axion inflation, the inflaton field $\phi$ couples to the abelian gauge field (which we call the electromagnetic field in the following) by means of the $\propto I(\phi)\,F_{\mu\nu}F_{\alpha\beta}\epsilon^{\mu\nu\alpha\beta}$ interaction term. Here $F_{\mu\nu}$ is the gauge-filed stress tensor and $I(\phi)$ is the axial coupling function which has to be a pseudoscalar in order to preserve the parity symmetry. The simplest and most popular choice for this function is the linear dependence on the pseudoscalar inflaton field, $I(\phi)\propto \phi$. The presence of nontrivial coupling to the inflaton breaks the conformal invariance of the Maxwell action which is the necessary condition for the generation of electromagnetic fields from quantum fluctuations \cite{Parker:1968}.
In axion inflation model, electromagnetic modes of only one circular polarization get enhanced due to interaction with the inflaton, which results in nonzero helicity of the produced field. 

Apart from the magnetic field, the electric one is also generated during inflation leading to two main consequences: (i)~it produces pairs of charged particles due to the Schwinger effect \cite{Sauter:1931,Heisenberg:1936,Schwinger:1951} and (ii)~nonzero scalar product $\boldsymbol{E}\cdot \boldsymbol{B}$ gives rise to the chiral asymmetry in fermionic sector because of the chiral anomaly \cite{Adler:1969gk,Bell:1969ts}. The Schwinger pair production during inflation was studied in Refs.~\cite{Domcke:2018,Sobol:2019,Domcke:2019,Kobayashi:2014,Froeb:2014,Bavarsad:2016,Stahl:2016a,Stahl:2016b,Hayashinaka:2016a,Hayashinaka:2016b,Sharma:2017,Bavarsad:2018,Geng:2018,Hayashinaka:2018,Hayashinaka:thesis,Giovannini:2018a,Banyeres:2018,Stahl:2018,Kitamoto:2018,Sobol:2018,Shtanov:2020,Tangarife:2017,Chua:2019,Shakeri:2019,Gorbar:2019,Sobol:2020Sch,Domcke:2021yuz}. In particular, it was shown that this phenomenon may significantly affect the generation of gauge fields and make a sizeable contribution to the Universe reheating. However, relatively little attention was payed to the chiral anomaly during axion inflation \cite{Domcke:2018,Domcke:2019}. To the best of our knowledge, the quantitative analysis of the chiral charge generation is still missing in the literature. Also, the backreaction of this chiral asymmetry on the gauge-field evolution due to the chiral magnetic effect has not been investigated yet. This provides the main motivation for the study in this paper.

An efficient approach to study the magnetogenesis during axion inflation which was proposed and applied quite recently \cite{Sobol:2019,Gorbar:2021} is the gradient expansion formalism. It deals with a set of bilinear electromagnetic functions in position space which are the vacuum expectation values of scalar products of the electric and magnetic field vectors with an arbitrary number of spatial curls. Its advantage compared to the conventional approach handling separate Fourier modes of the electromagnetic fields is that it automatically takes into account all relevant modes simultaneously and thus can be applied even in the presence of such complicated nonlinear phenomena as the backreaction and the Schwinger effect which couple all electromagnetic modes to each other. In this work we extend the gradient expansion formalism by taking into account the presence of chiral asymmetry in the fermionic sector (characterized by the chiral chemical potential $\mu_{5}$) and the additional term in the electric current induced by this chiral asymmetry -- the chiral magnetic effect current $\boldsymbol{j}_{\rm CME}=e^{2}/(2\pi^2)\mu_{5}\boldsymbol{B}$ (for review see Ref.~\cite{Kharzeev:2013ffa} and references therein).

The paper is organized as follows. The axion inflation model and the gradient expansion formalism for its description are discussed in Sec.\ref{sec-model}. The chiral magnetic effect is introduced and incorporated into the gradient expansion formalism in Sec.\ref{sec-chiral}. Numerical results for the chirality production in axion inflation are presented in Sec.\ref{sec-num} and summarized in Sec.\ref{sec-summary}.

\section{Magnetogenesis in axial-coupling model}
\label{sec-model}

In axion inflation, the coupling of the electromagnetic field $A_{\mu}$ to the pseudoscalar inflaton field $\phi$ breaks the conformal invariance of the
Maxwell action and, thus, is crucial for the electromagnetic field generation. The action of the axion inflation model is given by
\begin{equation}
\label{S}
S=\int d^{4}x\sqrt{-g} \left[\frac{1}{2}g^{\mu\nu}\partial_{\mu}\phi\,\partial_{\nu}\phi - V(\phi)-\frac{1}{4}F_{\mu\nu}F^{\mu\nu}-\frac{1}{4}I(\phi)F_{\mu\nu}\tilde{F}^{\mu\nu} +\mathcal{L}_{\rm f}(\psi,\,A_{\mu})\right],
\end{equation}
where $g={\rm det\,}g_{\mu\nu}$ is the determinant of the spacetime metric, $V(\phi)$ is the inflaton potential, $I(\phi)$ is the axial-coupling function, $F_{\mu\nu}=\partial_{\mu}A_{\nu}-\partial_{\nu}A_{\mu}$ is the gauge-field strength tensor, and its dual tensor is defined by
\begin{equation}
\label{F}
\tilde{F}^{\mu\nu}=\frac{1}{2\sqrt{-g}}\,\varepsilon^{\mu\nu\lambda\rho}F_{\lambda\rho},
\end{equation}
where $\varepsilon^{\mu\nu\lambda\rho}$ is the absolutely antisymmetric Levi-Civita symbol with $\varepsilon^{0123}=+1$. The last term in Eq.~(\ref{S}) describes the matter fields charged under the $U(1)$ gauge group and, therefore, coupled to the electromagnetic four-potential $A_{\mu}$. In this work, we consider a toy model in which only one massless charged fermionic field $\psi$ is present. Moreover, we will not consider the dynamics of this matter field in detail and only describe its observable quantities such as the electric current, energy density etc. 
	
The following Euler--Lagrange equations for the inflaton and electromagnetic fields are easily obtained from action (\ref{S}):
\begin{equation}
\label{L_E_1}
\frac{1}{\sqrt{-g}}\partial_{\mu} \left[\sqrt{-g}\, g^{\mu\nu}\,\partial_{\nu}\phi \right] + \frac{dV}{d\phi} + \frac{1}{4}\frac{dI}{d\phi}F_{\mu\nu}\tilde{F}^{\mu\nu}=0,
\end{equation}
\begin{equation}
\label{L_E_2}
\frac{1}{\sqrt{-g}}\partial_{\mu}\left[\sqrt{-g}\,F^{\mu\nu} \right]+ \frac{dI}{d\phi}\,\tilde{F}^{\mu\nu}\partial_{\mu}\phi=j^{\nu},
\end{equation}
where $j^{\nu}=-\partial \mathcal{L}_{\rm f}(\psi,\,A_{\mu})/\partial A_{\nu}$ is the electric four-current of fermions induced by the electromagnetic field.

In addition, the dual gauge field strength tensor satisfies the Bianchi identity
\begin{equation}
	\label{Bianchi}
	\frac{1}{\sqrt{-g}}\partial_{\mu}\left[\sqrt{-g}\,\tilde{F}^{\mu\nu} \right] = 0.
\end{equation}

The energy--momentum tensor in our model is given by
\begin{equation}
\label{T}
T_{\mu\nu}=
\partial_{\mu}\phi\,\partial_{\nu}\phi - g^{\lambda\rho}F_{\mu\lambda}F_{\nu\rho}-g_{\mu\nu}\left[\frac{1}{2}\partial_{\alpha}\phi\,\partial^{\alpha}\phi
-V(\phi)-\frac{1}{4}F_{\alpha\beta}F^{\alpha\beta}\right]+T_{\mu\nu}^{\rm f},
\end{equation}
where the last term defines the contribution of the charged fermionic field. The zero-zero component of $T_{\mu\nu}$ gives the energy density of the Universe.

Now, we restrict ourselves to the case of the spatially-flat FLRW metric and assume that the inflaton field $\phi$ is spatially homogeneous, i.e., only depends on time. It is convenient to use the Coulomb gauge for the electromagnetic four-potential $A_{\mu}=(0,\,-\boldsymbol{A})$. Then, the three-vectors of electric $\bm{E}$ and magnetic $\bm{B}$ fields, as measured by the comoving observer, are defined in the conventional way
\begin{equation}
\label{EB}
\bm{E}=-\frac{1}{a}\partial_{0}\bm{A}, \quad \bm{B}=\frac{1}{a^{2}}{\rm rot\,}\bm{A}.
\end{equation}
The gauge-field stress tensor and its dual tensor are expressed in terms of the electric and magnetic fields as follows:
\begin{equation}
F_{0i}=aE^{i}, \quad F_{ij}=-a^{2}\varepsilon_{ijk} B^{k}, \quad \tilde{F}_{0i}=aB^{i}, \quad \tilde{F}_{ij}=a^{2}\varepsilon_{ijk} E^{k}.
\end{equation}
Finally, we assume that the plasma of charged fermions is spatially homogeneous and quasineutral, i.e., $j^{0}=0$, while the electric current density is described by the Ohm's law
\begin{equation}
	j^{i}=\frac{1}{a}\sigma_{\rm f} E^{i}.
\end{equation}
It was shown in the literature that this is indeed the case for fermions produced by the Schwinger effect during inflation if the electromagnetic field is constant \cite{Hayashinaka:2016a,Domcke:2019}. In such a case the generalized conductivity $\sigma_{\rm f}$ depends only on the absolute values of the electric and magnetic fields. We will use the same form of the electric current assuming that the electromagnetic field changes adiabatically slowly. The applicability of this approximation was discussed in detail in Ref.\cite{Gorbar:2021}.

Then, the full system of equations in our model consists of the Friedmann equation for the Hubble parameter $H=\dot{a}/a$ determining the expansion rate of the Universe
\begin{equation}
\label{Friedmann}
H^{2}=\frac{\rho}{3 M_{\mathrm{p}}^{2}}=\frac{1}{3 M_{\mathrm{p}}^{2}}\left[\frac{1}{2}\dot{\phi}^{2}+V(\phi) + \frac{1}{2}\left\langle \bm{E}^{2}+\bm{B}^{2} \right\rangle +\rho_{\rm f}\right],
\end{equation}
the Klein-Gordon equation for the inflaton field
\begin{equation}
	\label{KGF}
	\ddot{\phi}+3H\dot{\phi}+V'(\phi)=I'(\phi)\left\langle \bm{E}\cdot\bm{B} \right\rangle,
\end{equation}
and Maxwell's equations for the electromagnetic field
\begin{equation}
	\label{Maxwell_1}
	\dot{\bm{E}}+2 H \bm{E}-\frac{1}{a} {\rm rot} \bm{B} + I'(\phi)\,\dot{\phi}\,\bm{B}+\sigma_{\rm f}\bm{E}=0,
\end{equation}
\begin{equation}
	\label{Maxwell_2}
	\dot{\bm{B}}+2 H \bm{B}+\frac{1}{a} {\rm rot} \bm{E}=0,
\end{equation}
\begin{equation}
	\label{Maxwell_3}
	{\rm div} \bm{E}=0, \qquad {\rm div} \bm{B}=0.
\end{equation}
In Eq.~(\ref{Friedmann}), $M_{\mathrm{p}}=(8\pi G)^{-1/2}=2.43\times 10^{18}\,\text{GeV}$ is the reduced Planck mass. Since we consider the electromagnetic field as a quantum field, the vacuum expectation value is taken when it appears in the classical equations of motion (\ref{Friedmann}) and (\ref{KGF}). They are denoted by the angle brackets.

In order to close the system of equations, we need to specify the conductivity. In the case of one massless Dirac fermion species with charge $e$, the conductivity induced by the Schwinger effect is constant and collinear electric and magnetic fields in de Sitter spacetime has the form \cite{Domcke:2019}
\begin{equation}
	\label{sigma}
	\sigma_{\mathrm{f}}=\frac{e^{3}}{6\pi^{2}}\frac{|B|}{H}{\rm coth}\Big(\frac{\pi|B|}{|E|}\Big),
\end{equation}
where $|E|\equiv\sqrt{\langle \boldsymbol{E}^{2}\rangle}$ and $|B|\equiv\sqrt{\langle \boldsymbol{B}^{2}\rangle}$. During axion inflation the generated gauge fields are helical and the vectors $\boldsymbol{E}$ and $\boldsymbol{B}$ are indeed nearly collinear. Expression~(\ref{sigma}) was derived in the strong-field regime, $|eE|\gg H^{2}$, which is the most important one for physical applications.

The energy of the electric field which is dissipated due to finite conductivity is transferred into charged fermions produced by the Schwinger effect. Then, the energy density of fermions satisfies the following equation which can be derived using the energy balance in the system:
\begin{equation}
	\label{rho-eq}
	\dot{\rho}_{\rm f}+4H\rho_{\rm f}=\sigma_{\rm f} \langle\boldsymbol{E}^{2}\rangle.
\end{equation}

For massless fermions, apart from the ordinary $U(1)$ gauge symmetry, at classical level there is also the global $U(1)$ chiral symmetry which ``rotates'' phases of right- and left-handed components of the spinor $\psi$ with opposite signs. According to Noether's theorem, it should correspond to a conserved current $j^{\mu}_{5}$ which is the difference of currents of right- and left-handed particles. However, in quantum field theory this symmetry is violated by renormalization of ultraviolet divergences. As a result, the corresponding Noether current is not conserved (the chiral anomaly, see Refs.~\cite{Adler:1969gk,Bell:1969ts}). In the expanding Universe, the chiral anomaly equation reads as
\begin{equation}
	\frac{1}{\sqrt{-g}}\partial_{\mu}(\sqrt{-g}j^{\mu}_{5})=-\frac{e^{2}}{8\pi^{2}}F_{\mu\nu}\tilde{F}^{\mu\nu}.
\end{equation}
Assuming again the homogeneous fermion distribution, we rewrite this equation in the form
\begin{equation}
	\label{anomaly-eq}
	\dot{n}_{5}+3Hn_{5}=\frac{e^{2}}{2\pi^{2}}\langle \boldsymbol{E}\cdot\boldsymbol{B}\rangle,
\end{equation}
where $n_{5}=j^{0}_{5}$ is the chiral charge density of fermions.\footnote{We do not take into account perturbative chirality-flipping processes which equilibrate the chiral asymmetry, since they are negligible at temperatures above $80\,$TeV \cite{Campbell:1992jd,Bodeker:2019ajh} which is true in our model, see Sec.~\ref{sec-num}.} Since in the axion inflation the nonzero value of $\langle \bm{E}\cdot\bm{B} \rangle$ is generated, the fermion chirality is not conserved. At the beginning of inflation there were no gauge fields and fermions (thus, the initial value of chiral asymmetry was equal to zero). Then, the nonconservation of chiral charge implies that the nonzero $n_{5}$ will be generated during axion inflation.


In order to study this process numerically, one needs to solve the coupled system of Friedmann (\ref{Friedmann}), Klein-Gordon (\ref{KGF}), and Maxwell equations (\ref{Maxwell_1})--(\ref{Maxwell_3}) together with the anomaly equation (\ref{anomaly-eq}). However, the Maxwell equations (\ref{Maxwell_1})--(\ref{Maxwell_3}) describe the evolution of quantum electric and magnetic fields; therefore, one has to deal with mode functions of these fields in Fourier space. In the presence of backreaction or the Schwinger effect all modes become coupled to each other that makes numerical calculations very demanding. To overcome this problem, we utilize the gradient expansion formalism developed in Ref.~\cite{Gorbar:2021}. It operates with vacuum expectation values of bilinear electromagnetic functions in coordinate space that include all physically relevant modes at once. These functions are defined as follows:
\begin{equation}
\label{E_n}
\mathcal{E}^{(n)}=\frac{1}{a^{n}}\left\langle \bm{E}\cdot {\rm rot}^{n} \bm{E}  \right\rangle,
\end{equation}
\begin{equation}
\label{G_n}
\mathcal{G}^{(n)}=-\frac{1}{a^{n}}\left\langle \bm{E}\cdot {\rm rot}^{n} \bm{B}  \right\rangle,
\end{equation}	
\begin{equation}
\label{B_n}
\mathcal{B}^{(n)}=\frac{1}{a^{n}}\left\langle \bm{B}\cdot {\rm rot}^{n} \bm{B}  \right\rangle.
\end{equation}
Equations of motion for these quantities can be derived from the Maxwell's equations (\ref{Maxwell_1}) and (\ref{Maxwell_2}) and have the form~\cite{Sobol:2019,Gorbar:2021}:
\begin{equation}
\label{dot_E_n}
\dot{\mathcal{E}}^{(n)} + [(n+4)H+2\sigma_{\rm f}]\,	\mathcal{E}^{(n)} - 2I'(\phi)\dot{\phi}\,\mathcal{G}^{(n)} +2\mathcal{G}^{(n+1)}=[\dot{\mathcal{E}}^{(n)}]_{b},
\end{equation}
\begin{equation}
\label{dot_G_n}
\dot{\mathcal{G}}^{(n)} +[(n+4)H+\sigma_{\rm f}]\, \mathcal{G}^{(n)}-\mathcal{E}^{(n+1)}+\mathcal{B}^{(n+1)} - I'(\phi)\dot{\phi}\,\mathcal{B}^{(n)}=[\dot{\mathcal{G}}^{(n)}]_{b},
\end{equation}
\begin{equation}
\label{dot_B_n}
\dot{\mathcal{B}}^{(n)} + (n+4)H\,	\mathcal{B}^{(n)}-2\mathcal{G}^{(n+1)}=[\dot{\mathcal{B}}^{(n)}]_{b}.
\end{equation}
where $[\dot{\mathcal{E}}^{(n)}]_{b}$, $[\dot{\mathcal{G}}^{(n)}]_{b}$, and $[\dot{\mathcal{B}}^{(n)}]_{b}$ are the boundary terms.

The origin of these terms is explained by the following. All possible modes of quantum gauge field fluctuations are always present in physical vacuum. Due to a quasiexponential expansion of the Universe during inflation the wavelength of some modes of electromagnetic field become larger than the Hubble horizon (more precisely, they become at some point tachyonically unstable and stop oscillating). Such long-wavelength modes can be treated as the Fourier modes of a classical electromagnetic field \cite{Lyth:2008}. Modes deep inside the horizon oscillate in time without significant change of their amplitude. Their total energy is infinite and its contribution to the electromagnetic energy density should be excluded. The number of modes that cross the horizon and  become physically relevant constantly grows during inflation. This leads to an additional time dependence of electromagnetic quantities which can be described by means of boundary terms in the equations of motion.

The boundary terms for the system of equations~(\ref{dot_E_n})--(\ref{dot_B_n}) were derived for the first time in Ref.~\cite{Sobol:2019} and were calculated more accurately in Ref.~\cite{Gorbar:2021}, including the impact of the Schwinger conductivity $\sigma_{\rm f}$. Considering the electromagnetic modes with circular polarization $\lambda=\pm$ one could obtain
\begin{equation}
\label{E_p_d}
[\dot{\mathcal{E}}^{(n)}]_{b}=\frac{d \ln k_{\mathrm{h}}(t)}{d t}\frac{\Delta(t)}{4\pi^{2}}\left(\frac{k_{\mathrm{h}}(t)}{a(t)}\right)^{n+4}\sum_{\lambda=\pm 1}\lambda^{n} E_{\lambda}(\xi(t),s(t)),
\end{equation}
\begin{equation}
\label{G_p_d}
[\dot{\mathcal{G}}^{(n)}]_{b}=\frac{d \ln k_{\mathrm{h}}(t)}{d t}\frac{\Delta(t)}{4\pi^{2}}\left(\frac{k_{\mathrm{h}}(t)}{a(t)}\right)^{n+4}\sum_{\lambda=\pm 1}\lambda^{n+1}G_{\lambda}(\xi(t),s(t)),
\end{equation}
\begin{equation}
\label{B_p_d}
[\dot{\mathcal{B}}^{(n)}]_{b}=\frac{d \ln k_{\mathrm{h}}(t)}{d t}\frac{\Delta(t)}{4\pi^{2}}\left(\frac{k_{\mathrm{h}}(t)}{a(t)}\right)^{n+4}\sum_{\lambda=\pm 1}\lambda^{n}B_{\lambda}(\xi(t),s(t)),
\end{equation}
where $k_{\mathrm{h}}(t)$ is the momentum of the mode that crosses the horizon at time $t$
\begin{equation}
\label{k-h}
k_{\mathrm{h}}(t)=\underset{t'\leq t}{\rm max}\Big\{a(t')H(t')\big[|\xi(t')|+\sqrt{\xi^{2}(t')+s^{2}(t')+s(t')}\big]\Big\},
\end{equation}
\begin{equation}
\label{xi-s}
\xi(t)=\frac{dI}{d\phi}\frac{\dot{\phi}}{2H}, \qquad s(t)=\frac{\sigma_{\rm f}(t)}{2H},
\end{equation}
the parameter $\Delta$ is the exponential of the integrated conductivity,
\begin{equation}
\label{Delta-parameter}
\Delta(t)\equiv \exp\Bigg(-\int\limits_{-\infty}^{t}\sigma_{\rm f}(t')dt'\Bigg),
\end{equation}
and
\begin{equation}
\label{E-lambda}
E_{\lambda}(\xi,s)=\frac{e^{\pi\lambda \xi}}{r^{2}(\xi,s)} \left|\left(i r(\xi,s)-i\lambda \xi-s\right)W_{-i\lambda\xi,\frac{1}{2}+s}(-2i r(\xi,s))+W_{1-i\lambda\xi,\frac{1}{2}+s}(-2i r(\xi,s))\right|^{2},
\end{equation}
\begin{equation}
\label{G-lambda}
G_{\lambda}(\xi,s)=
\frac{e^{\pi\lambda \xi}}{r(\xi,s)} \left\{\Re e\left[W_{i\lambda \xi,\frac{1}{2}+s}(2i r(\xi,s)) W_{1-i\lambda\xi,\frac{1}{2}+s}(-2i r(\xi,s))\right]-s\left|W_{-i\lambda\xi,\frac{1}{2}+s}(-2i r(\xi,s)) \right|^{2}\right\},
\end{equation}
\begin{equation}
\label{B-lambda}
    \qquad B_{\lambda}(\xi,s)=e^{\pi\lambda \xi}\,\left|W_{-i\lambda\xi,\frac{1}{2}+s}(-2i r(\xi,s)) \right|^{2}
\end{equation}
with $r(\xi,s)=|\xi|+\sqrt{\xi^{2}+s+s^{2}}$ and the Whittaker function $W_{\kappa,\mu}(y)$. For large $\xi$, these expressions can be expanded into series in inverse powers of $r(\xi,s)$, which is convenient for numerical computations. 

Equations (\ref{dot_E_n})--(\ref{dot_B_n}) form an infinite chain because the equation of motion for the $n$th order function always contains at least one function with the $(n+1)$th power of the curl. However, this chain can be truncated at some large order $n_{max}$ due to a simple observation. For a large power $n\gg 1$ of the spatial curl the main contribution to $\mathcal{E}^{(n)}$, $\mathcal{B}^{(n)}$, $\mathcal{G}^{(n)}$ is made by modes with the largest possible momentum which is the momentum of mode crossing the horizon at the moment of time under consideration $k_{h}(t)$. Then, one can write
\begin{equation}
	\mathcal{E}^{(n_{max}+1)}\approx\Big(\frac{k_{h}(t)}{a(t)}\Big)^{2}\mathcal{E}^{(n_{max}-1)}
\end{equation}
and similar relations for $\mathcal{B}^{(n_{max}+1)}$, $\mathcal{G}^{(n_{max}+1)}$ which allow us to truncate the chain at some order $n_{max}$. We have chosen to relate the quantities with orders differing by 2 because they have the same symmetry properties under the spatial inversion. After the truncation we get a closed system of ordinary differential equations that describe the self-consistent evolution of classical observables in the form of quadratic functions of the electric and magnetic fields with an arbitrary power of the curl.

\section{Chiral magnetic effect}
\label{sec-chiral}

In the previous section, we obtained a coupled system describing the generation of electromagnetic fields and charged fermions during axion inflation with a special attention to the production of chiral asymmetry in the fermionic sector. However, the latter may also have an impact on electromagnetic fields due to the \textit{chiral magnetic effect} phenomenon. It leads to nonzero electric current proportional to the magnetic field with proportionality coefficient related to the chiral asymmetry (for review see Ref.~\cite{Kharzeev:2013ffa} and references therein):
\begin{equation}
	\label{cme}
	\boldsymbol{j}_{\rm CME}=\frac{e^{2}}{2\pi^{2}}\mu_{5} \boldsymbol{B}.
\end{equation}
Here, $\mu_{5}$ is the chiral chemical potential which is a conjugated quantity to $N_{5}=\int n_{5}\,d^{3}\boldsymbol{x}$ in equilibrium thermodynamics. In other words, it characterizes the average energy that one needs to spend in order to increase the chiral charge of the system $N_{5}$ by unity. 

In order to determine $\mu_{5}$, we assume that fermions produced by the Schwinger effect rapidly thermalize and can be characterized by the equilibrium Fermi-Dirac distribution function with some temperature $T$. Although such an approximation is not well justified far from the end of inflation when the number of produced fermions is very low, it seems to be reasonable during the final part of axion inflation where the generated electromagnetic field is strong and the Schwinger pair creation effectively produces many fermions. In any case, such an equilibrium approximation will allow us to estimate the impact of the chiral magnetic effect on the electromagnetic field generation.

Since at the beginning of inflation there were no fermions and the Schwinger effect can produce them only in particle-antiparticle pairs, the plasma is quasineutral and the electric chemical potential $\mu=0$. Then, momentum distribution functions for right- and left-handed particles have the form:
\begin{equation}
	\label{distr-chiral}
	f_{R/L}(p)=\frac{1}{\exp\big(\frac{p\mp \mu_{5}}{T}\big)+1},
\end{equation}
i.e., the chemical potential for right-handed particles equals to $\mu_{5}$ while for left-handed ones $-\mu_{5}$. Distribution functions for the corresponding antiparticles $\bar{f}_{R/L}$ can be obtained from Eq.~(\ref{distr-chiral}) by the replacement $\mu_{5}\to -\mu_{5}$. Then, the chiral charge and the total energy density of fermions can be expressed in terms of $T$ and $\mu_{5}$ as follows:
\begin{equation}
	\label{n5}
	n_{5}=\int\frac{d^{3}\boldsymbol{p}}{(2\pi)^{3}}\left[f_{R}(p)-f_{L}(p)-\bar{f}_{R}(p)+\bar{f}_{L}(p)\right]=\frac{1}{3}\mu_{5}T^{2}+\frac{1}{3\pi^{2}}\mu_{5}^{3},
\end{equation}
\begin{equation}
	\label{rho-f}
	\rho_{\rm f}=\int\frac{d^{3}\boldsymbol{p}}{(2\pi)^{3}}\,p\,\left[f_{R}(p)+f_{L}(p)+\bar{f}_{R}(p)+\bar{f}_{L}(p)\right]=\frac{7\pi^{2}}{60}T^{4}+\frac{1}{2}\mu_{5}^{2}T^{2}+\frac{1}{4\pi^{2}}\mu_{5}^{4}.
\end{equation}
Here we used the fact that the antiparticle of a right-handed particle has left-handed chirality and vice versa. We would like to note that Eqs.~(\ref{n5}) and (\ref{rho-f}) give exact results of integration of the Fermi-Dirac functions without any assumptions about the relative magnitude of $\mu_{5}$ and $T$. These nonlinear relations allow to use interchangeably two sets of variables, either $n_{5}$ and $\rho_{\rm f}$ or $\mu_{5}$ and $T$. In order to take into account the chiral magnetic effect, the latter set of variables is more convenient since the current (\ref{cme}) depends on $\mu_{5}$.

Using Eqs.~(\ref{rho-eq}) and (\ref{anomaly-eq}) together with relations (\ref{n5}) and (\ref{rho-f}), we derive the following equations of motion for $T$ and $\mu_{5}$:
\begin{equation}
	\label{dot-T}
	\dot{T}+HT=\frac{1}{\pi^{2} T\,D(T,\,\mu_{5})}\left[\Big(\pi^{2}T^{2}+3\mu_{5}^{2}\Big)\sigma_{\rm f} \mathcal{E}^{(0)}+3\mu_{5}\Big(\pi^{2}T^{2}+\mu_{5}^{2}\Big)\frac{e^{2}}{2\pi^{2}}\mathcal{G}^{(0)}\right],
\end{equation}
\begin{equation}
	\label{dot-mu5}
	\dot{\mu}_{5}+H\mu_{5}=-\frac{1}{D(T,\,\mu_{5})}\left[2\mu_{5}\sigma_{\rm f} \mathcal{E}^{(0)}+\Big(\frac{7\pi^{2}}{5}T^{2}+3\mu_{5}^{2}\Big) \frac{e^{2}}{2\pi^{2}}\mathcal{G}^{(0)}\right],
\end{equation}
where 
\begin{equation}
	D(T,\,\mu_{5})=\frac{7\pi^{2}}{15}T^{4}+\frac{2}{5}\mu_{5}^{2}T^{2}+\frac{1}{\pi^{2}}\mu_{5}^{4}=\frac{7\pi^{2}}{15}\Big(T^{2}+\frac{3}{7\pi^{2}}\mu_{5}^{2}\Big)^{2}+\frac{32}{35\pi^{2}}\mu_{5}^{4}>0.
\end{equation}

In order to determine the impact of the chiral magnetic effect on the gauge-field production, we add current (\ref{cme}) to the left-hand side of the Maxwell equation (\ref{Maxwell_1}). This current is similar to the forth term in Eq.~(\ref{Maxwell_1}) since it is also proportional to $\boldsymbol{B}$. Therefore, effectively we have to replace 
\begin{equation}
	I'(\phi)\dot{\phi}\ \rightarrow\ I'(\phi)\dot{\phi}+\frac{e^{2}}{2\pi^{2}}\mu_{5}
\end{equation}
everywhere in the equations of motion for electromagnetic fields, in particular, in Eqs.~(\ref{dot_E_n})--(\ref{dot_B_n}) which then take the form
\begin{equation}
	\label{dot_E_n-2}
	\dot{\mathcal{E}}^{(n)} + [(n+4)H+2\sigma_{\rm f}]\,	\mathcal{E}^{(n)} - 2\Big(I'(\phi)\dot{\phi}+\frac{e^{2}}{2\pi^{2}}\mu_{5}\Big)\,\mathcal{G}^{(n)} +2\mathcal{G}^{(n+1)}=[\dot{\mathcal{E}}^{(n)}]_{b},
\end{equation}
\begin{equation}
	\label{dot_G_n-2}
	\dot{\mathcal{G}}^{(n)} +[(n+4)H+\sigma_{\rm f}]\, \mathcal{G}^{(n)}-\mathcal{E}^{(n+1)}+\mathcal{B}^{(n+1)} - \Big(I'(\phi)\dot{\phi}+\frac{e^{2}}{2\pi^{2}}\mu_{5}\Big)\,\mathcal{B}^{(n)}=[\dot{\mathcal{G}}^{(n)}]_{b},
\end{equation}
\begin{equation}
	\label{dot_B_n-2}
	\dot{\mathcal{B}}^{(n)} + (n+4)H\,	\mathcal{B}^{(n)}-2\mathcal{G}^{(n+1)}=[\dot{\mathcal{B}}^{(n)}]_{b}.
\end{equation}
Note that the parameter $\xi$ in Eqs.~(\ref{E_p_d})--(\ref{B-lambda}) is replaced by 
\begin{equation}
	\xi_{\rm eff}=\xi+\frac{e^{2}}{4\pi^{2}}\frac{\mu_{5}}{H}
\end{equation}
and thus $\mu_{5}$ modifies also the boundary terms.

\section{Numerical results}
\label{sec-num}

Let us now specify the inflationary model and consider chirality generation in it numerically. We use the simplest quadratic inflaton potential
\begin{equation}
	\label{potential-infl}
	V(\phi)=\frac{m^{2}\phi^{2}}{2},
\end{equation}
which well describes the behavior of a wide class of inflaton potentials close to their minima. This is important since the most intensive generation of the electromagnetic fields and charged fermions occurs close to the end of inflation when the inflaton indeed approaches the minimum of its potential. For definiteness, in numerical analysis we take $m=6\times 10^{-6}\,M_{\mathrm{p}}$.

The axial coupling constant $I(\phi)$ is taken in a linear form 
\begin{equation}
	\label{coupling-f}
	I(\phi)=\beta\frac{\phi}{M_{\rm p}}
\end{equation}
with a dimensionless parameter $\beta$. This is the typical expression for the coupling function for the axion-like inflaton field. For numerical analysis, we use the range of values of $\beta=10-25$. For smaller values the generated fields are very weak while larger values would lead to the generation of big non-Gaussianities in the primordial scalar power spectrum \cite{Guzzetti:2016}.

Initial conditions for the inflaton field and its time derivative are chosen from the requirement that the inflation stage lasts at least 60 $e$-foldings, i.e., $\phi(0)\approx15.5\,M_{\rm P}$ and $\dot{\phi}(0)=-\sqrt{2/3}MM_{\rm P}$. The latter expression was derived in the slow-roll approximation for potential (\ref{potential-infl}). Note that only the last 10--15 $e$-foldings are important for magnetogenesis and fermions production; however, the initial conditions should be specified well before this moment. Zero initial values for all electromagnetic quantities $\mathcal{E}^{(n)}$, $\mathcal{B}^{(n)}$, and $\mathcal{G}^{(n)}$ as well as for the fermion energy density $\rho_{\rm f}$ and chiral charge $n_{5}$ were assumed.

One important comment on the introduction of the chiral chemical potential should be made. Nonlinear relations (\ref{n5}) and (\ref{rho-f}), generally speaking, do not allow to determine $\mu_{5}$ and $T$ for arbitrary values of $\rho_{\rm f}$ and $n_{5}$. In fact, assuming that Eqs.~(\ref{n5}) and (\ref{rho-f}) are satisfied, the following relation is valid:
\begin{equation}
	\frac{3\sqrt{\pi}\,n_{5}}{2\sqrt{2}\,\rho_{\rm f}^{3/4}}=\frac{1+\tfrac{\pi^{2}T^{2}}{\mu_{5}^{2}}}{\Big(1+\tfrac{2\pi^{2}T^{2}}{\mu_{5}^{2}}+\tfrac{7\pi^{4}T^{4}}{15\mu_{5}^{4}}\Big)^{3/4}}.
\end{equation}
One can easily check that the right-hand side of this equation lies in the range $(0,\,1)$ for all possible values of $\mu_{5}$ and $T$. Consequently, if for given values of $\rho_{\rm f}$ and $n_{5}$, the combination $\frac{3\sqrt{\pi}\,n_{5}}{2\sqrt{2}\,\rho_{\rm f}^{3/4}}>1$,  Eqs.~(\ref{n5}), (\ref{rho-f}) cannot be inverted and $\mu_{5}$ and $T$ cannot be identified. In this situation we conclude that fermions cannot be characterized by the thermal Fermi-Dirac distribution function. This indeed happens in our numerical analysis at very early times when the particle production is very slow. It is quite natural since for low densities of particles their thermalization is very unlikely in the exponentially expanding Universe.

In order to overcome this difficulty, at 60 $e$-foldings to the end of inflation we impose initial conditions for $\rho_{\rm f}$ and $n_{5}$ rather than for $\mu_{5}$ and $T$. (Of course, in this case we are not able to take into account the chiral magnetic effect in the Maxwell equations; however, for low fermion densities this effect can be neglected.) We numerically solve our system of equations for the electromagnetic functions (\ref{dot_E_n})--(\ref{dot_B_n}) together with Eqs.~(\ref{rho-eq}) and (\ref{anomaly-eq}) for $\rho_{\rm f}$ and $n_{5}$ until some moment of time when the condition $\frac{3\sqrt{\pi}\,n_{5}}{2\sqrt{2}\,\rho_{\rm f}^{3/4}}<1$ is satisfied. At this moment of time we are able to invert Eqs.~(\ref{n5}), (\ref{rho-f}) and determine the corresponding values of $\mu_{5}$ and $T$. They are taken as initial conditions for the second stage where we solve Eqs.~(\ref{dot-T}) and (\ref{dot-mu5}) for $\mu_{5}$ and $T$ together with Eqs.~(\ref{dot_E_n-2})--(\ref{dot_B_n-2}) for the electromagnetic bilinear function which now take into account the chiral magnetic effect.

\begin{figure}[ht!]
	\centering
	\includegraphics[height=3.9cm]{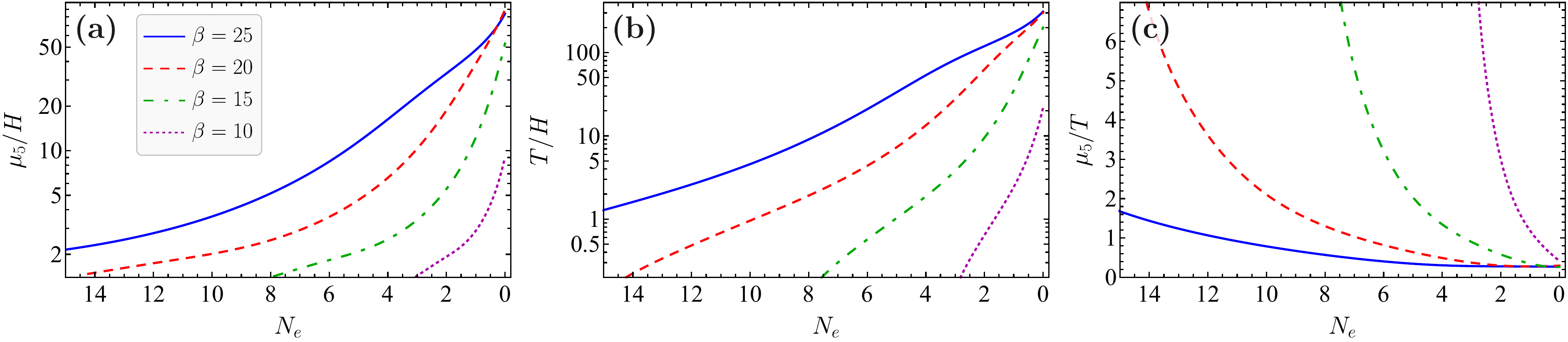}
	\caption{The chiral chemical potential in Hubble units $\mu_{5}/H$ [panel~(a)], the effective temperature in Hubble units $T/H$ [panel~(b)], and their ratio $\mu_{5}/T$ [panel~(c)] in dependence of the number of $e$-foldings counted from the end of inflation $N_{e}$ for four different values of the axial coupling parameter: $\beta=10$ (purple dotted lines), $\beta=15$ (green dashed-dotted lines), $\beta=20$ (red dashed lines), and $\beta=25$ (blue solid lines).}
	\label{fig-1}
\end{figure}

We present the results for $\mu_{5}$ and $T$ generated in our model as functions of the number of $e$-foldings to the end of inflation in Fig.~\ref{fig-1}. Panel~(a) shows the chiral chemical potential in the Hubble units, panel~(b) shows the temperature in the Hubble units, while panel~(c) represents their ratio. The lines of different color and style correspond to four different values of the axial coupling constant: $\beta=10$ (purple dotted lines), $\beta=15$ (green dashed-dotted lines), $\beta=20$ (red dashed lines), and $\beta=25$ (blue solid lines). 

First of all, we would like to note that both temperature and chiral chemical potential grow in time which corresponds to the fact that more fermions are produced by the Schwinger effect and more fermions change their chirality due to the chiral anomaly. However, the ratio $\mu_{5}/T$ decreases in time meaning that the relative fraction of chiral imbalance $n_{5}=n_{R}-n_{L}$ compared to the total fermion number $n_{\rm tot}=n_{R}+n_{L}$ decreases in time. This can be explained by the fact that close to the end of inflation the electromagnetic field becomes very strong and the Schwinger effect produces a lot of fermions. Being initially equally distributed between the chiralities, they flip their chirality due to the chiral anomaly which results in the chiral imbalance. However, the latter process is less intensive since it is proportional to the second power of the electromagnetic field while the Schwinger pair production term in Eq.~(\ref{rho-eq}) is proportional to the third power of the electromagnetic field. Thus, more particles are produced in total than particles which change their chirality due to the chiral anomaly. Thus, we can conclude that the chiral imbalance $n_{5}/n_{\rm tot}$ becomes less as time passes.

Figure~\ref{fig-1}(c) implies that the ratio of the final chiral chemical potential to temperature at the end of inflation weakly depends on the axial coupling parameter $\beta$ and equals approximately to $\mu_{5}/T\sim 0.27-0.3$. This can be translated into the chiral asymmetry by the following considerations. For $\mu_{5}/T\ll 1$, the total number density of right- and left-handed fermions equals to
\begin{equation}
    n_{\rm tot}=n_{R}+n_{L}=\int\frac{d^{3}\boldsymbol{p}}{(2\pi)^{3}}\,\left[f_{R}(p)+f_{L}(p)+\bar{f}_{R}(p)+\bar{f}_{L}(p)\right]= \frac{3\zeta(3)}{\pi^{2}}T^{3}+\text{O}(\mu_{5}^{2}T),
\end{equation}
where $\zeta(3)$ is the Riemann zeta function. Then, taking into account that in the same approximation $n_{5}=n_{R}-n_{L}\approx \mu_{5}T^{2}/3$, we obtain
\begin{equation}
\eta=\frac{n_{5}}{n_{\rm tot}}=\frac{n_{R}-n_{L}}{n_{R}+n_{L}}\approx \frac{\pi^{2}}{9\zeta(3)}\,\frac{\mu_{5}}{T} \approx 0.91 \frac{\mu_{5}}{T}.
\end{equation}
Thus we conclude that an excess of one chirality with respect to the other can reach $25\%$. Thus, the axion inflation is indeed an efficient chirality generator.
	
\begin{figure}[ht!]
	\centering
	\includegraphics[height=4.5cm]{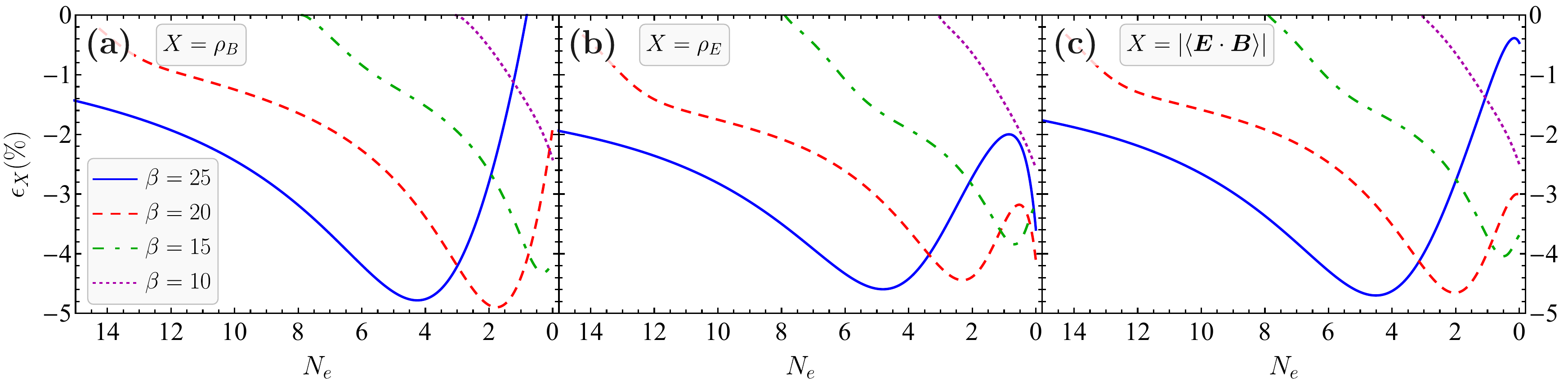}
	\caption{The relative change of the magnetic energy density $\rho_{B}$ [panel~(a)], electric energy density $\rho_{E}$ [panel~(b)], and
		Chern-Pontryagin density $|\langle \boldsymbol{E}\cdot \boldsymbol{B}|$, caused by the presence of the chiral magnetic effect,
		in dependence of the number of $e$-foldings counted from the end of inflation $N_{e}$
		for four different values of the axial coupling parameter: $\beta=10$ (purple dotted lines), $\beta=15$ (green dashed-dotted lines),
		$\beta=20$ (red dashed lines), and $\beta=25$ (blue solid lines).}
	\label{fig-2}
\end{figure}

Now let us discuss the impact of the chiral magnetic effect on the evolution of the electromagnetic field. For this purpose, we solved Eqs.~(\ref{dot_E_n-2})--(\ref{dot_B_n-2}) in the presence of the chiral magnetic effect and compared the results with the solutions of Eqs.~(\ref{dot_E_n})--(\ref{dot_B_n}) where this effect is switched off. The relative deviation
\begin{equation}
	\epsilon_{X}=\frac{X-X_{\rm ref}}{X_{\rm ref}}\times 100\%,
\end{equation}
where $X=\{\rho_{E},\,\rho_{B},\,|\langle\boldsymbol{E}\cdot\boldsymbol{B}\rangle|\}$ and $X_{\rm ref}$ is the solution without chiral magnetic effect. We show this relative deviations in Fig.~\ref{fig-2} again for four values of $\beta$ considered above.

Analyzing this plot, we conclude that the impact of the chiral magnetic effect is rather weak since the deviation is of order $1-5\%$. Qualitatively, it leads to the decrease of the electromagnetic field (note the negative values of $\epsilon_{X}$). This is natural since the electromagnetic field is the source of chiral asymmetry (due to the chiral anomaly) while the chiral magnetic effect represents the backreaction of this asymmetry on the electromagnetic field. Unless this backreaction works to decrease the electromagnetic field, one gets an instability in the system. 
As we discussed above, the axial coupling with the inflaton field and the chiral magnetic effect have similar structure. However, during inflation the latter is much less important than the former and a general tendency to increase the electromagnetic field is observed.

\section{Conclusion}
\label{sec-summary}

In this work, we studied the generation of electromagnetic fields and charged fermions during axion inflation taking into account the backreaction of generated fields on the inflaton evolution, the Schwinger effect, and the chirality production due to the chiral anomaly. Moreover, we took into account the chiral magnetic effect and analyzed its impact on the magnetogenesis. For this purpose, we employed the gradient expansion formalism previously proposed by three of us in Refs.~\cite{Sobol:2019,Gorbar:2021}. Since it operates with bilinear electromagnetic functions in the position space, it takes into account all physically relevant modes at once and thus allows us to treat the highly nonlinear phenomena listed above.

Qualitatively, the picture of the chirality production is the following. Far from the end of inflation, the electromagnetic field is weak and the Schwinger pair production is very slow (see, e.g., Ref.~\cite{Gorbar:2021}). As a result, the number density of produced particles is small, and we cannot assume that they are thermalized. Indeed, the estimates in Ref.~\cite{Domcke:2019} for exact de Sitter spacetime and constant electromagnetic field confirm that thermalization does not occur. However, close to the end of inflation, the inflaton rolls faster and, therefore, the electromagnetic field becomes very strong (for sufficiently large $\beta$ the electromagnetic energy density can be comparable with that of the inflaton, see Ref.~\cite{Sobol:2019,Gorbar:2021}). It is natural to assume that the charged particles which are very intensively produced due to the Schwinger effect quickly achieve the thermal distribution. In this assumption, we introduced the effective temperature $T$ and chiral chemical potential $\mu_{5}$ which characterizes the chiral asymmetry of fermions. Although both quantities grow in time due to particle and chirality production, the ratio $\mu_{5}/T$ is a decreasing function of time. Indeed, at earlier times, there is a small number density of fermions and the chiral anomaly transfers most them into one chiral state resulting in $\mu_{5}\gg T$. At later times, however, more and more fermions are created by the Schwinger effect (which produces the same amount of particles with both chiralities), and the chiral anomaly cannot transfer all of them into one chirality. As a result, the ratio $\mu_{5}/T$ decreases.

We performed our numerical analysis in a simple inflationary model with quadratic potential $V(\phi)=m^{2}\phi^{2}/2$ and for a linear axial coupling function $I(\phi)=\beta\phi/M_{\rm P}$ with a dimensionless parameter $\beta$. We showed that despite the fact that $\mu_{5}/T$ rapidly decreases in time during the last few $e$-foldings of inflation, the final residual value of this quantity at the end of inflation weakly depends on the coupling parameter $\beta$ and is of order 0.27--0.3. Numerically, this corresponds to 25--27\% of chiral imbalance in the fermion number density. Thus, we conclude that axion inflation is an efficient chirality generator.

Nonzero value of the chemical potential $\mu_{5}$ leads to a new contribution to the electric current -- the chiral magnetic effect $\boldsymbol{j}\propto \mu_{5}\boldsymbol{B}$ which backreacts on the electromagnetic field evolution. We analyzed numerically this impact and concluded that such a backreaction leads to only a few percent change in the electric and magnetic energy densities. Thus, the impact of the chiral magnetic effect on the gauge field evolution is less important than their interaction with the inflaton. We would expect, however, that this phenomenon may be very important in the postiflationary evolution of the magnetic field since at that time the interaction with the inflaton will not play any role. Since our gradient expansion formalism is applicable only during inflation, some other methods are needed in order to investigate the postinflationary evolution in this system. We keep this problem for further studies.

Let us note once more that the results obtained in this work are based on the assumption that starting from a certain moment a few $e$-foldings prior to the end of inflation the fermions become thermalized and can be characterized by the temperature and chemical potential. It was crucial to estimate the value of $\mu_{5}$ in order to analyze its impact on the electromagnetic field, because the chiral magnetic effect depends on $\mu_{5}$. However, since this impact is very weak, one can neglect the chiral magnetic effect during inflation and study the chirality production without any assumptions about the fermion distribution. The most efficient tool to perform such an analysis is the kinetic approach. We plan to address this issue in the further studies.

\begin{acknowledgments}	
	The work of E.~V.~G., A.~I.~M., and I.~V.~R. was supported by the National Research Foundation of Ukraine Project No.~2020.02/0062. The work of O.~O.~S. was supported by the ERC-AdG-2015 grant 694896. The work of S.~I.~V. was supported by the Swiss National Science Foundation Grant No.~SCOPE IZSEZ0 206908.	
\end{acknowledgments}

 \end{document}